\documentclass[twocolumn]{aastex62}
\usepackage{graphicx}
\usepackage{color}
\usepackage{amsmath}

\begin{document}

\title{AGN Disks Harden the Mass Distribution of Stellar-mass Binary Black Hole Mergers}

\author{Y. Yang}
\affiliation{Department of Physics, University of Florida, PO Box 118440, Gainesville, FL 32611-8440, USA}
\author{I. Bartos$^*$}
\affiliation{Department of Physics, University of Florida, PO Box 118440, Gainesville, FL 32611-8440, USA}
\email{$^*$Email: imrebartos@phys.ufl.edu}
\author{Z. Haiman}
\affiliation{Department of Astronomy, Columbia University, 550 W 120th St, New York, NY 10027, USA}
\author{B. Kocsis}
\affiliation{Institute of Physics, E\"otv\"os University, P\'azm\'any P. s. 1/A, Budapest, 1117, Hungary}
\author{Z. M\'arka}
\affiliation{Department of Physics, Columbia University, 550 W 120th St., New York, NY 10027, USA}
\author{N.C. Stone}
\affiliation{Department of Physics, Columbia University, 550 W 120th St., New York, NY 10027, USA}
\author{S. M\'arka}
\affiliation{Department of Physics, Columbia University, 550 W 120th St., New York, NY 10027, USA}

\begin{abstract}
The growing number of stellar-mass binary black hole mergers discovered by Advanced LIGO and Advanced Virgo are starting to constrain the binaries' origin and environment. However, we still lack sufficiently accurate modeling  of binary formation channels to obtain strong constraints, or to identify sub-populations. One promising formation mechanism that could result in different black hole properties is binaries merging within the accretion disks of Active Galactic Nuclei (AGN). Here we show that the black holes' orbital alignment with the AGN disks preferentially selects heavier black holes. We carry out Monte Carlo simulations of orbital alignment with AGN disks, and find that AGNs harden the initial black hole mass function. Assuming an initial power law mass distribution $M_{\rm bh}^{-\beta}$, we find that the power law index changes by $\Delta \beta\sim1.3$, resulting in a more top-heavy population of merging black holes. This change is independent of the mass of, and accretion rate onto, the supermassive black hole in the center of the AGN. Our simulations predict an AGN-assisted merger rate of $\sim4$\,Gpc$^{-3}$yr$^{-1}$. With its hardened mass spectra, the AGN channel could be responsible for $10-50$\% of gravitational-wave detections.
\end{abstract}

\section{Introduction}

Stellar-mass binary black holes (BBHs) can form in multiple distinct astrophysical sites and processes. Formation channels include isolated binary stellar systems in which both stars end their lives as black holes \citep{2002ApJ...572..407B,2008ApJ...676.1162S,2014LRR....17....3P}, and dynamical formation in which the black holes become gravitationally bound following a chance encounter \citep{2000ApJ...528L..17P,2007PhRvD..76f1504O,2009ApJ...692..917M,2013LRR....16....4B}. With the increasing rate of BBH mergers discovered by Advanced LIGO \citep{2015CQGra..32g4001L} and Advanced Virgo \citep{TheVirgo:2014hva}, our prospects to probe these formation mechanisms is rapidly improving \citep{2018arXiv181112907T}.

Active galactic nuclei (AGNs) represent a promising site for the dynamical formation and/or merger of BBHs. Galactic centers harbor a large, dense population of stellar-mass black holes (BHs) due to mass segregation \citep{1977ApJ...216..883B,2006ApJ...648..411K,2009MNRAS.395.2127O,2018Natur.556...70H}. In active galaxies the orbit of some of these BHs will align with the accretion disk of the central supermassive black hole (SMBH) due to gas damping \citep{1991MNRAS.250..505S,2007MNRAS.374..515L,2012MNRAS.425..460M}. BHs within the disk can migrate due to angular momentum exchange with the disk, towards so-called migration traps \citep{2012MNRAS.425..460M}. As BHs accumulate near these traps, they can form BBH systems \citep{2016ApJ...819L..17B,2018arXiv180702859S}. Additionally, a significant fraction of the BHs may already enter the disk as binaries \citep{2017ApJ...835..165B}, while some BH binaries could be born within the disk itself \citep{2017MNRAS.464..946S}. Once a binary enters or is formed within the SMBH accretion disk, it rapidly ($\lesssim1$\,Myr) merges due to gaseous torques \citep{2017ApJ...835..165B,2017MNRAS.464..946S}.
\begin{figure*}
\begin{center}
\includegraphics[width=0.65\textwidth]{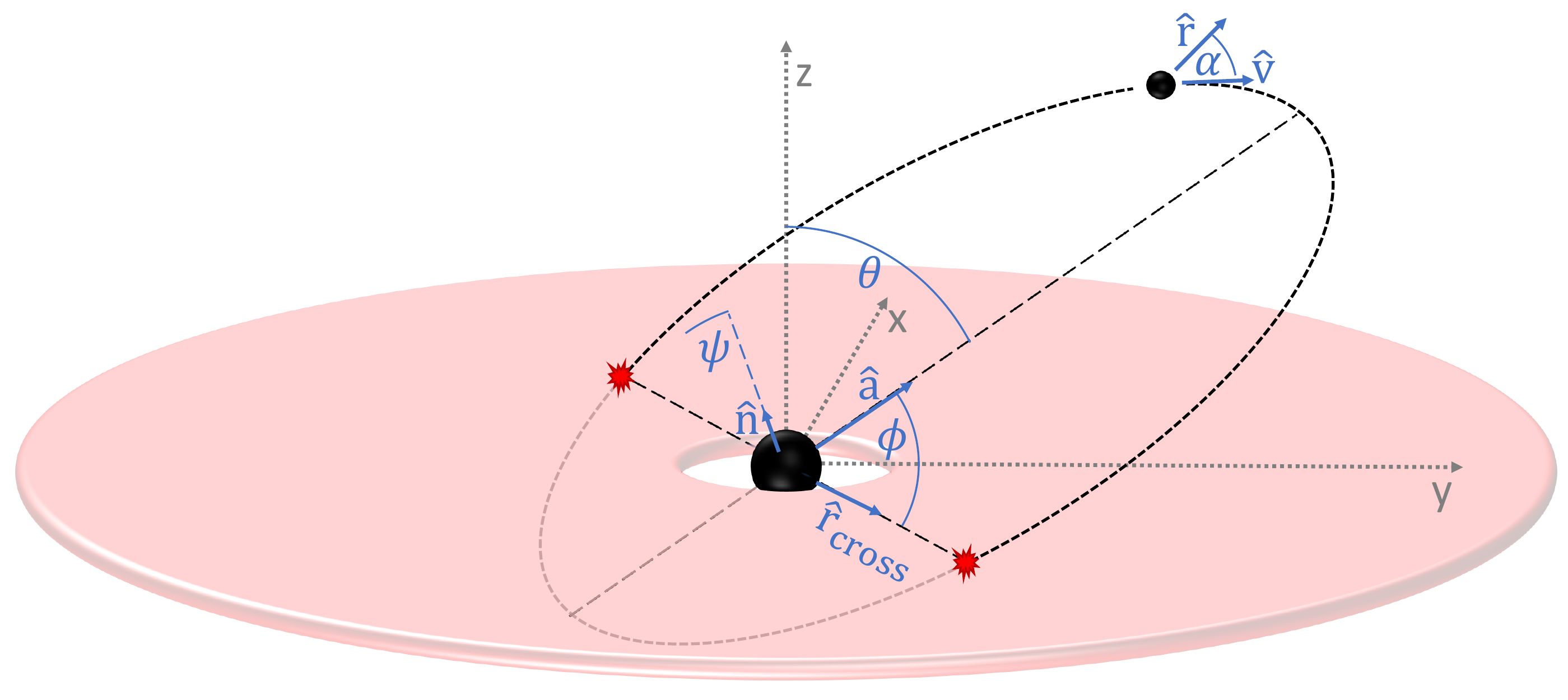}%
\end{center}
\caption{\label{fig:illustration} Sketch of the AGN disk and the stellar mass BH orbit indicating the parameters used in the calculation. }
\end{figure*}

An interesting feature of BBH mergers in AGNs is the possibility of detectable electromagnetic emission. As the BBH system is surrounded by a dense gaseous environment, the BHs can form mini-accretion disks and may reach super-Eddington accretion rates \citep{2017ApJ...835..165B,2017MNRAS.464..946S}. Emission from the mini-disks must be very bright, and possibly relativistically beamed, in order to be detected, as the resulting radiation needs to outshine the AGN in which the BBH was embedded in. BBH mergers detected by Advanced LIGO \citep{2015CQGra..32g4001L} and Advanced Virgo \citep{2015CQGra..32b4001A} have been regularly followed-up by electromagnetic and neutrino observations \citep{2016ApJ...826L..13A,2016PhRvD..93l2010A,2017PhRvD..96b2005A}, increasing the chance that any electromagnetic emission from these mergers will be found. Already, there have been claims of the coincident detection of gamma-ray bursts with BBH mergers \citep{2016ApJ...826L...6C,2017ApJ...847L..20V}. More coincident observations will help clarify whether these associations were astrophysical.

Apart from electromagnetic counterparts, determining the formation channel of BBH mergers relies on the mergers' reconstructed properties. An important distinguishing feature can be the host galaxy of the binaries. While the localization of BBH mergers with Earth-based interferometers is mostly inadequate to constrain their origin \citep{2016arXiv161201471C}, a set of the best-localized mergers can be sufficient to establish an association with certain rare host galaxy types \citep{2017NatCo...8..831B}. This might be particularly helpful with BBH mergers in AGNs, as AGNs represent only a small fraction of galaxies \citep{2005AJ....129.1795H}.

A powerful differentiator between formation scenarios is the mass and spin distribution of the merging BHs \citep{2019MNRAS.482.2991A}. BBH systems from isolated stellar binaries are expected to favor close to equal masses \citep{2012ApJ...759...52D,2016MNRAS.458.2634M}, and to have BH spins aligned with the orbital axis \citep{2017PhRvL.119a1101O,2018PhRvD..98h4036G,2017MNRAS.467.2146K,2019MNRAS.484.4216R,2016ApJ...816...65A,2018ApJ...856..140H}. Dynamically formed binaries typically have unequal BH masses, a top-heavy mass distribution compared to the initial mass function \citep{2016ApJ...824L..12O,2016PhRvD..93h4029R}, and randomly oriented spins compared to their orbital axis. In particular, spin orientation has been explored as a means to differentiate between formation scenarios \citep{2017CQGra..34cLT01V,2017Natur.548..426F}. The spin components parallel to the orbital axis for BBH mergers detected by Advanced LIGO are mostly consistent with zero, which suggest either no spin alignment, which is the expected outcome of dynamical formation \citep{2017Natur.548..426F}. It could also be explained by low BH spins at merger \citep{2017ApJ...842..111H}.

The expected properties of BBH mergers in AGNs are currently poorly understood. \cite{2018ApJ...866...66M} considered the possible range of astrophysical parameters that determine the merger rate, and found that a very broad range of rates, within $\Gamma=10^{-3}-10^{4}\,$Gpc$^{-3}$yr$^{-1}$, is possible. They also showed via a simplified model that the migration of the black holes within the disk, and gas accretion can modify the spins and masses of merging black holes. In particular, they pointed out that if the post-merger BH stays within the disk then the mass distribution of merging black holes will shift to significantly heavier objects due to repeated mergers.

In this paper we explore the role of the alignment of the orbital plane of BHs around SMBHs with AGN disk planes on the parameters and rate of BBH mergers. These dependencies have not been explored previously even though it has major influence on BBH mergers. We considered a realistic distribution of SMBH masses, accretion rates onto the SMBH, and multiple initial BH mass distributions. We focused on the scenario in which singleton BHs ground down to alignment with the AGN disk and form binaries in migration traps within the disk.

The paper is organized as follows. In Section \ref{sec:method} we present our analytical framework to calculate the alignment rate of black holes as a function of the properties of the BH, the SMBH and the SMBH accretion disk. In Section \ref{sec:results} we present our findings on BHs in the AGN disk and on the properties of binary mergers within the disk. In Section \ref{sec:starformation} we briefly describe an alternative source of stellar-mass BHs in AGN disks from star formation within the disk. We conclude in Section \ref{sec:conclusion}.

\section{Method}
\label{sec:method}

\subsection{AGN disk}
We consider a geometrically thin optically thick, radiatively efficient and steady-state accretion disk \citep{1973A&A....24..337S}, which is expected in AGNs.  
We assume the disk has viscosity parameter $\alpha=0.1$, and radiation efficiency $\epsilon=L_{\bullet,Edd}/\dot{M}_{\bullet,Edd}c^2$, where $L_{\bullet,Edd}$ and $\dot{M}_{\bullet,Edd}$ are the Eddington luminosity and Eddington accretion rate of a SMBH with mass $M_{\bullet}$. We further assume the accretion rate $\dot{m}_{\bullet}=\dot{M}_{\bullet}/\dot{M}_{\bullet,Edd}$ is constant during the lifetime $\tau_{\rm agn}$ of an AGN. These parameters determine the disk surface density $\Sigma(r)$, scale height $H(r)$, isothermal sound speed $c_{\rm s}(r)$, and midplane temperature $T(r)$, which are dependent on the distance $r$ from the SMBH. To derive these quantities we follow the model of \cite{2003MNRAS.341..501S}, assuming that viscosity in the disk is proportional to the total pressure:
\begin{align}
\sigma T_{\rm eff}^4&=\frac{3}{8\pi}\dot{M}'\Omega^2\\
T^4&=\left(\frac{3}{8}\tau+\frac{1}{4\tau}+\frac{1}{2}\right)T_{\rm eff}^4\\
\tau&=\frac{\kappa\Sigma}{2}\\
\beta^bc_{\rm s}^2\Sigma&=\frac{\dot{M}'\Omega}{3\pi\alpha}\\
p_{\rm rad}&=\frac{\tau\sigma}{2c}T_{\rm eff}^4\\
p_{\rm gas}&=\frac{\rho kT}{m}\\
\beta&=\frac{p_{\rm gas}}{p_{\rm gas}+p_{\rm rad}}\\
\Sigma&=2\rho H\\
c_{\rm s}&=\rho H=\sqrt{\frac{p_{\rm gas}+p_{\rm rad}}{\rho}}
\end{align}
Where $T_{\rm eff}(r)$ is the black body temperature of the disc at r. $\dot{M}'=\dot{M}_{\bullet}\sqrt{1-r_{\rm  min}/r}$, $r_{\rm min}$ is the inner radius of the disc, we set $r_{\rm min}=3r_{\rm s}$ here, where $r_{\rm s}=2GM/c^2$. $\Omega$ is the  Keplerian angular velocity of the The disc, which is $\sqrt{GM/r^3}$. $\tau$ is the optical depth at midplane. We assume the mean molecular mass $m=0.62m_{\rm H}$. $\sigma$ is the Stefan-Boltzmann constant and $k$ is Boltzmann constant. In order to solve these equations, we still need to provide the opacity $\kappa$. We approximate the opacity with (c.f. Fig. 1 in \citealt{2005ApJ...630..167T}):\\
\begin{equation}\label{}
\kappa(\rho,T)=\left\{
\begin{aligned}
&\kappa_0T^2 \quad&T<100K\\
&\kappa_1 \quad&T<1000K\\
&\frac{\kappa_{\rm m}-\kappa_{\rm 1}}{500K}(T-1000K)+\kappa_1 \quad&T<1500K\\
&\kappa_{\rm m}+(\kappa^{-1}_{\rm H^-}+(\kappa_{\rm e}+\kappa_{\rm K})^{-1})^{-1}
\end{aligned}
\right.
\end{equation}
Where $\kappa_0=2.4\times10^{-4}$\,g$^{-1}$cm$^2$K$^{-2}$, $\kappa_1=\kappa_0(100$K$)^2=2.4$\,g$^{-1}$cm$^2$. $\kappa_m$ is the molecular opacity, which is about $0.1Z$, where $Z$ indicates mass fraction of heavy elements. We adopt $Z=0.01$ in our simulations \citep{1996ApJ...464..943I}. $\kappa_{H^-}$ is the opacity due to  the negative hydrogen ion: \begin{equation}
\kappa_{H^-}\approx 2\,Z^{0.5}\rho_{-11}^{0.5}T_{4}^{7.7}\,\mbox{g}^{-1}\mbox{cm}^2,
\end{equation}
where $\rho_{-11}\equiv\rho/(10^{-11}\,\mbox{g\,cm}^{-3})$ and $T_{4} = T / (10^4\,\mbox{K})$. The electron scattering opacity $\kappa_e$ can be written as 
\begin{multline}
\kappa_e = 0.2\,(1+X)\left(1+2.7\times10^{-8}\frac{\rho_{-11}}{T_{4}^{2}}\right)^{-1} \\
\times\left[1+\left(\frac{T_{4}}{4.5\times10^4}\right)^{0.86}\right]^{-1}\,\mbox{g}^{-1}\mbox{cm}^2
\end{multline}
$X$ is the mass fraction of hydrogen. The Kramers opacity $\kappa_K$ is due to free-free, bound-free, and bound-bound electronic transitions: 
\begin{equation}
\kappa_{\rm K}\approx 4\,(1+X)(Z+0.001)\rho_{-11} T_{4}^{-3.5}\,\mbox{g}^{-1}\mbox{cm}^2.
\end{equation}
The equations above are valid where the Toomre Q-parameter $Q=C_{\rm s}\Omega/(\pi G\Sigma)\gtrsim1$. Beyond that region, out to 0.1\,pc, we assume that the disk self-regulates, which we take into account by fixing the Q-parameter at 1 and replace Eq. (1) with (see Eq. 15 of \citealt{2003MNRAS.341..501S}):
\begin{equation}\label{}
\rho=\frac{\Omega^2}{2\pi G}
\end{equation}

\subsection{Orbital variation due to AGN disk crossing}

Consider a Cartesian coordinate system with the SMBH in its center, in which the AGN disk lies in the XY plane, and the major axis of the orbit of the stellar mass BH orbiting the SMBH is in the YZ plane (see Fig. \ref{fig:illustration}). Let $\hat{\bf{a}}=(0,\sin{\theta},\cos{\theta})$ be the unit vector pointing in the direction of the major axis of the BH's orbit in Cartesian coordinates, with $\theta$ being the angle is the orbit's inclination angle compared to the AGN disk plane. Let $a$ be the semi-major axis, $e$ the eccentricity and $\hat{\bf{n}}$ is the normal vector of the BH's orbital plane. Let $\psi\equiv\pi/2+\hat{\bf{n}}\hat{\bf{x}}$ the angle between $\hat{\bf{n}}$ and YZ plane, where $\hat{\bf{x}}$ is the unit vector along the x axis\footnote{Here we adopt the coordinates $(\theta,\psi,\phi)$ that are the most suitable for the derivation of the results, even though these are not standard Keplerian orbital elements. They can be converted to Keplerian coordinates as follows. Longitude of the ascending mode: $\arccos[\cos\theta\cos\psi/(\sin^2\psi\sin^2\theta+\cos^2\theta)]$. Argument of periapsis: $\phi$. Inclination: $\sin\theta\sin\psi$.}. We have
\begin{equation}\label{}
\hat{\bf{n}}=(\sin{\psi},-\cos{\theta}\cos{\psi},\sin{\theta}\cos{\psi})
\end{equation}
Let $\phi$ be the angle between the major axis of the BH orbit and the intersection of the orbital plane and the AGN disk. We find that
\begin{equation}\label{}
\cos{\phi}=\frac{\sin{\psi}\sin{\theta}}{\sqrt{{\sin^2{\psi}}{\sin^2{\theta}}+\cos^2{\theta}}}
\end{equation}
The BH orbit crosses the AGN disk plane at two points. The distance of these crossing points from the focal point of the BH orbit are
\begin{equation}\label{}
r_{\rm cross}=\frac{a(1-e^2)}{1\mp e\cos{\phi}}
\end{equation}
where $-$($+$) corresponds to the farther(closer) point. The normal vectors pointing from the focal point to these two crossing points are
\begin{equation}
{\bf{\hat{r}_{cross}}}=\frac{\pm\left(\cos{\theta}\cos{\psi},\,\sin{\psi},\,0\right)}{\sqrt{{\sin^2{\psi}}{\sin^2{\theta}}+\cos^2{\theta}}}
\end{equation}
The energy of the BH orbit is 
\begin{equation}
E=-\frac{GM_{\bullet}M_{\rm bh}}{2a},
\end{equation}
where $M_{\bullet}$ is the SMBH mass and $M_{\rm bh}$ is the BH mass. Thus, the speed of the BH at the crossing points is 
\begin{equation}
v_{\rm bh}=\sqrt{GM_{\bullet}}\sqrt{\frac{2}{r}-\frac{1}{a}},
\end{equation}
\newline
where $r$ is the BH-SMBH distance.

Let $\alpha$ be the angle between the BH velocity and radius vectors, so the unit vector in the direction of velocity is
\begin{equation}\label{}
{\bf{\hat{v}_{bh}}}=\mp\cos{\alpha}\hat{r}+\sin{\alpha}\hat{r}\times\vec{n}
\end{equation}
Therefore, the $z$ component of the BH's velocity is 
\begin{equation}
v_{\rm bh,z}=v_{\rm bh}\sin{\alpha}\sqrt{{\sin^2{\psi}}{\sin^2{\theta}}+\cos^2{\theta}}
\end{equation}
The time of crossing is $t_{\rm cross}=2H/v_{\rm z}$, where H is the scale height and we used the fact that $v_{\rm z}$ does not change significantly in a single crossing. The velocity of the gas in the disk is 
\begin{equation}
{\bf {v}_{gas}}=\left(\frac{GM_{\bullet}}{r}\right)^{1/2}\hat{\bf{r}}\times\hat{\bf{z}}\,.
\end{equation}
The relative speed of the gas in the disk and BH is 
\begin{equation}
\Delta v=\sqrt{v_{\rm bh}^2+{v}_{\rm gas}^2-2v_{\rm bh}{v}_{\rm gas}\sin{\alpha}\sin{\theta}\cos{\psi}}\,.
\end{equation}
The mass accreted during the crossing is 
\begin{equation}
\Delta M_{\rm cross}=\Delta vt_{\rm cross}r_{\rm BHL}^2\pi\Sigma/(2H)\,,
\label{eq:mcross}
\end{equation} 
where $r_{\rm BHL}=2GM_{\rm bh}/(\Delta v^2+c_s^2)$ is the BH's Bondi-Hoyle-Lyttleton radius and $\Sigma$ is the surface density of the AGN disk. For Eq. \ref{eq:mcross} we also made use of the fact that $r_{\rm BHL}\ll H$. We define a dimensionless factor $\lambda \equiv \Delta M_{\rm cross}/M_{\rm bh}$. 
The change of velocity and angular momentum of the BH after a crossing are
\begin{align}
\bf{\Delta {v_{bh}}}&=-\lambda(\bf{v_{bh}}-\bf{v_{gas}})=-\lambda \bf{\Delta v}\\
\bf{\Delta J}&=-\lambda \bf{r}\times \bf{\Delta v} 
\end{align}
The crossing slightly changes most orbital parameters. Denoting the new parameters with $()^\prime$, we obtain
\begin{align}
{\bf v^\prime}&={\bf v_{bh}}+{\bf \Delta v_{bh}} \\
a'&=\left(\frac{2}{r}-\frac{{\bf v^\prime}^2}{GM_{\bullet}}\right)^{-1} \\
e'&=\sqrt{1-({\bf J}+{\bf \Delta J})^2/a'} \\
\phi'&=\arccos{\frac{1-a'(1-e'^2)/r}{e'}} \\
{\bf n}^\prime&=\frac{{\bf v}^\prime\times{\bf r}}{|{\bf v}^\prime\times{\bf r}|} \\
\hat{\bf a}^\prime&=\pm(\cos{\phi^\prime}\hat{\bf r}+\sin{\phi^\prime}\hat{\bf n}^\prime\times\hat{\bf r}) \\
\theta^\prime&=\arccos(\hat{\bf a}^\prime\cdot\hat{\bf z})=\arccos(\pm\sin{\phi^\prime}({\bf n}^\prime\times\hat{\bf r})\cdot\hat{\bf z}) \\
\mbox{$\psi^\prime$}&=\left\{ \begin{array}{rl}
\mbox{$\arcsin{\frac{(\hat{\bf a}^\prime\times\hat{\bf z})\cdot{\bf n}^\prime}{\sin{\theta^\prime}}}$}  & \mbox{if ${\bf n}^\prime\cdot\hat{\bf z}\geq0$} \\\\
\mbox{$\pi-\arcsin{\frac{(\hat{\bf a}^\prime\times\hat{\bf z})\cdot{\bf n}^\prime}{\sin{\theta^\prime}}}$} & \mbox{if ${\bf n}^\prime\cdot\hat{\bf z}<0$}\\
\end{array} \right.
\end{align}
These updated values above allow for the update of orbital parameters after each crossing. Repeating this procedure, we obtain the orbital evolution.\\\\

\subsection{Monte Carlo simulations}

We carried out Monte Carlo simulations to obtain the distribution of BH masses and mass ratios in BBH mergers within AGN disks, and the timescale of orbital alignment. We assumed that the orbits of singleton BHs align with the AGN disk, in which they form binaries near migration traps. Following the alignment, we consider the random interaction of two BHs, for simplicity without further interactions of the merged objects. This could occur, for example, if the merger produces a sufficiently large natal kick to move the object out of the disk. Additional interactions will result in heavier BHs within mergers. We assume that all BHs whose orbit aligns with the disk merge once, within a negligible time compared to the lifetime of the AGN disk. We neglect the mass increase of the black holes due to accretion from the AGN disk. We used a sample size of $10^4$ for a given parameter combination, except for the case of $M_\bullet=10^8$\,M$_\odot$ for which we used a sample size of 2000.

We considered a BH {\it initial} mass function $dN/dM_{\rm bh}\propto M_{\rm bh}^{-\beta}$ with BH masses $M_{\rm bh}\in[5\mbox{M}_{\odot},50\mbox{M}_{\odot}]$. The mass distribution of black holes is currently poorly understood. Black hole masses from Galactic X-ray binaries suggests a soft power law spectrum with $\beta\sim5$ \citep{2015MNRAS.446.1213K,2011ApJ...741..103F}, while LIGO/Virgo's observations indicate an index of $\beta = 0-3$ \citep{2018arXiv181112940T}. Here we adopted the index of the stellar initial mass function, $\beta=2.35$, as out fiducial model, and also considered $\beta=2$ and 3 to examine the $\beta$-dependence of our results.

As this initial BH population undergoes mass segregation \citep{1977ApJ...216..883B,2019MNRAS.484.3279P}, its mass distribution will vary with distance from the SMBH. We take into account this mass segregation in the spatial distribution of BHs. Following \cite{2009MNRAS.395.2127O}, we adopt an initial distribution as a function of the orbit's semi-major axis \citep{2018ApJ...860....5G}
\begin{equation}
\frac{dn}{da}\propto a^{-3/2-0.5M_{\rm bh}/M_{\rm max}}\,,
\end{equation}
with $M_{\rm max}=50$\,M$_{\odot}$. We see that higher-mass BHs will typically be closer to the SMBH. We considered the maximal semi-major axis of interest to be the radius of influence of the SMBH $R_{\rm inf}=1.2M_{6}^{1/2}$\,pc, where $M_{6}=M_{\bullet}/10^6$M$_{\odot}$. We further adopted a thermal eccentricity distribution of $n(e)=2e$ (see \citealt{2009ApJ...692.1075G}). Finally, we assumed that the AGN disk becomes inhomogeneous (clumpy) beyond a radius $R_{\rm disk}=0.1M_{6}^{1/2}$\,pc, once the disk's self-gravity becomes non-negligible \citep{2009ApJ...700.1952H}. The nature of BH-disk interactions in such inhomogeneous (clumpy) disks is unclear. As massive stars begin forming in the clumping disk, they generate pressure support through winds, radiation pressure and supernova explosions, which can keep the disk vertically supported and allow some gas to continue to accrete through into the inner, stable regions of the disk. Such inhomogeneous (clumpy) regions could still produce sufficient drag to align BH orbits with the disk. This may not reduce, on average, the orbital alignment timescale, but may broaden its distribution (for a fixed initial orbit). Nevertheless, due to the uncertainties in this picture, we conservatively assume that no interaction takes place in this inhomogeneous (clumpy) region of the disk.

Another source of inhomogeneities in the accretion disk is the presence of annular gaps associated with massive embedded objects \citep{2011PhRvD..84b4032K}.  While stellar-mass objects are unlikely to open these gaps at distances $r \gtrsim 0.01 ~{\rm pc}$ \citep{2017MNRAS.464..946S}, it is theoretically possible for embedded black holes to grow, through accretion, up to the isolation mass, which corresponds to an intermediate-mass black hole \citep{2004ApJ...608..108G,2012MNRAS.425..460M}.  The presence of such a gap would dramatically reduce the drag on any inclined orbits that pass through it, but we neglect this effect because the small geometric size of these gaps means that only a minor fraction of inclined orbits will interact with them at any given time.


Given the initial parameters of an orbit, we can calculate the orbit's alignment time $\tau_{\rm align}$ with the AGN disk, which we define as $1-\hat{\bf n}\cdot\hat{\bf z}<10^{-3}$. We carried out simulations assuming a fiducial AGN lifetime of $\tau_{\rm agn}=10^7$\,yr \citep{2004cbhg.symp..169M} and check which BHs aligned with the disk. We note that the lifetime of lower-luminosity AGNs is believed to be longer \citep{2005ApJ...632...81H,2005ApJ...630..716H}. To leading order, the alignment timescale for a stellar-mass BH with orbital period $t_{\rm orb}$ is $\tau_{\rm align} \sim t_{\rm orb} \lambda^{-1} /2$ \citep{1991MNRAS.250..505S,2017ApJ...835..165B}.

\section{Results}
\label{sec:results}

\subsection{Black hole mass function}

\begin{figure}
   \centering  
   \includegraphics[width=0.47\textwidth]{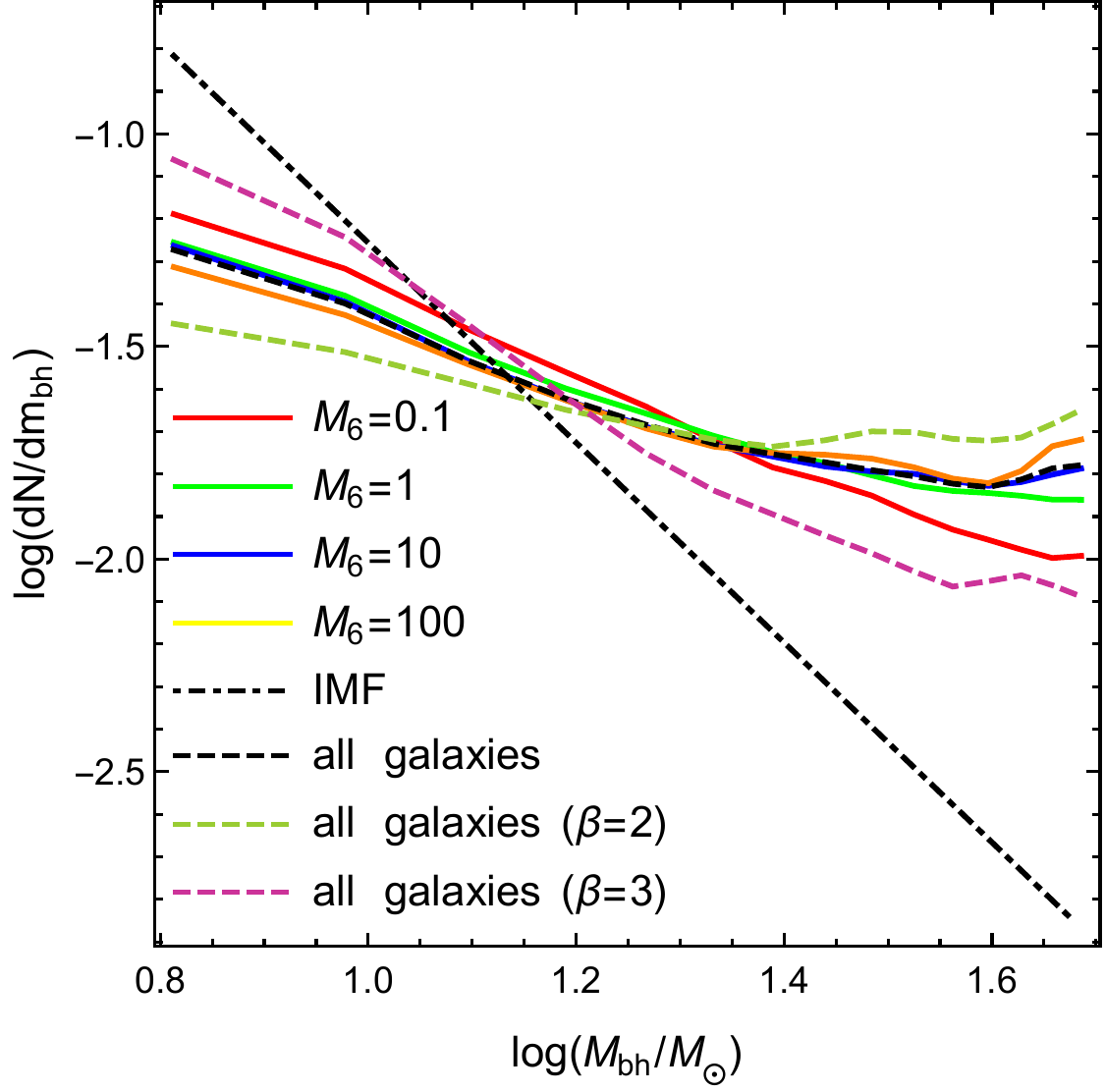}
   \caption{Simulated BH mass distributions for BBH mergers in AGN disks for different SMBH masses, and for all AGNs combined (see legend), assuming $\beta=2.35$. For comparison, we also show the obtained distribution for all galaxies for $\beta=2$ and $\beta=3$ (see legend), and the BH initial mass function (IMF) for $\beta=2$ (see legend). Both primary and secondary masses are included (separately) in the distribution. We see for all cases that the change is practically identical.}
\label{fig:massfunction}   
\end{figure}

We calculated the mass distribution of BHs that align with the AGN disk for a several initial mass function indices $\beta$ and SMBH masses. We found that after alignment the BH mass distribution still follows a power law, but the mass function index in the AGN disk changes from the initial value by $\Delta\beta\approx-1.3$. We found this change to be independent of both the initial $\beta$ index and the SMBH mass. We show the simulated mass distributions for different cases in Fig. \ref{fig:massfunction}.

There are three reasons that explain the flattening of the mass function. The first is mass segregation: since heavier BHs are closer to the SMBH, more of them cross the AGN disk plane within $R_{\rm disk}$ where the disk is continuous.  The second is alignment rate: heavier BHs align with the AGN disk more quickly, therefore they have larger chance to finish the evolution within the AGN lifetime. Third, heavier BHs accrete more matter while crossing the disk since $\Delta M_{\rm cross}\propto M_{\rm bh}^2$, making them slow down more during each crossing.

\subsection{Binary mass ratio}

Assuming that binaries are formed within the AGN disk from the random combination of captive BHs in the disk, we calculated the mass ratio $q\equiv M_1/M_2$ for BBH mergers, with BH masses $M_1\leq M_2$. We found that the distribution of $q$ peaks around $q\approx0.2$, with almost no mergers having lower $q<0.2$ and with a shallow distribution for $q>0.2$.  The lowest allowed mass ratio in our model is $q=0.1$. This means that the typical mass ratio for our AGN-merger model is low compared to mergers from isolated BH binaries \citep{2012ApJ...759...52D,2016MNRAS.458.2634M}. The mass ratio distribution we obtain is on average lower than what would be expected from the random pairing of BHs drawn from the initial mass function due to the harder mass spectrum in the AGN.

\begin{figure}
   \centering  
   \includegraphics[width=0.47\textwidth]{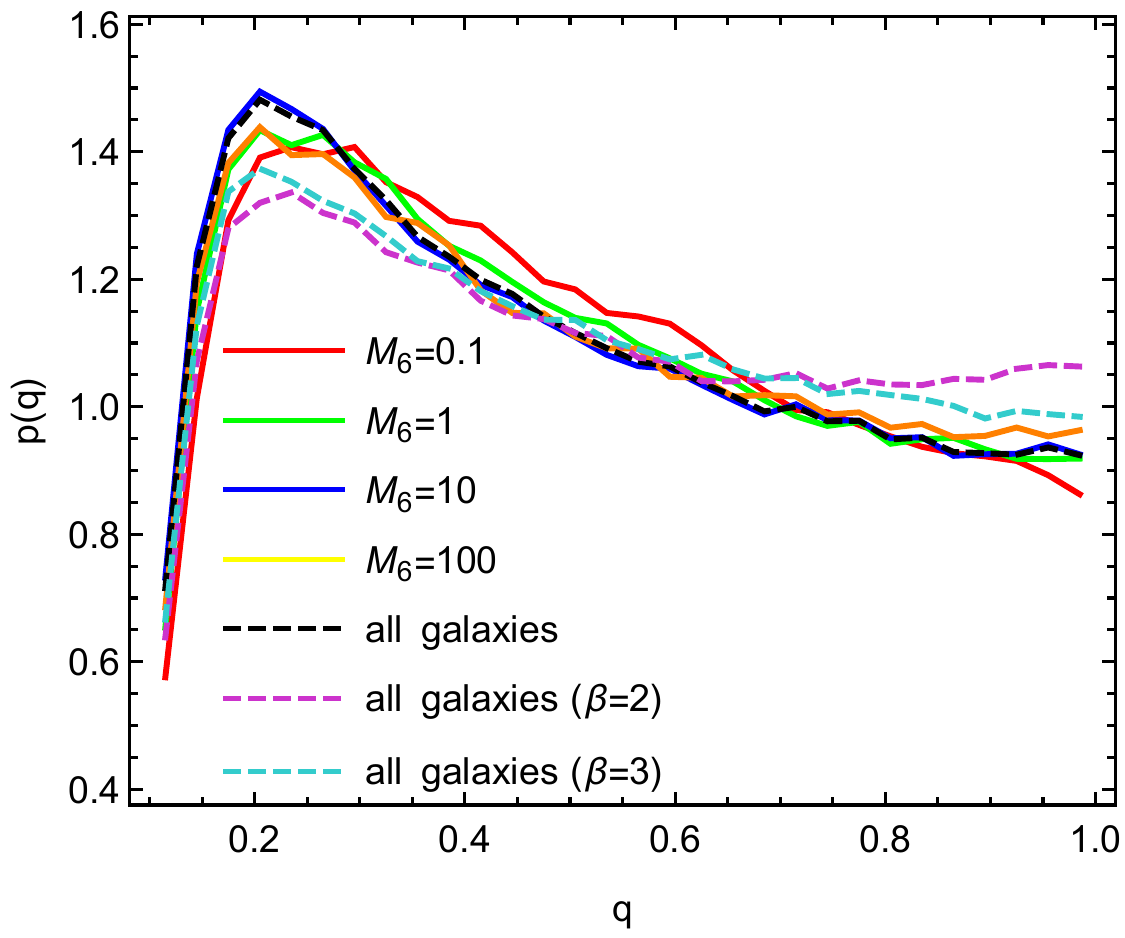}
   \caption{Probability density $p(q)$ of BBH mass ratio $q$ for mergers in AGN disks for different SMBH masses, and for all AGNs combined (see legend), assuming $\beta=2.35$. For comparison, we also show the obtained distribution for all galaxies for $\beta=2$ and $\beta=3$ (see legend).}
\label{fig:q}
\end{figure}

\begin{figure}
   \centering  
   \includegraphics[width=0.47\textwidth]{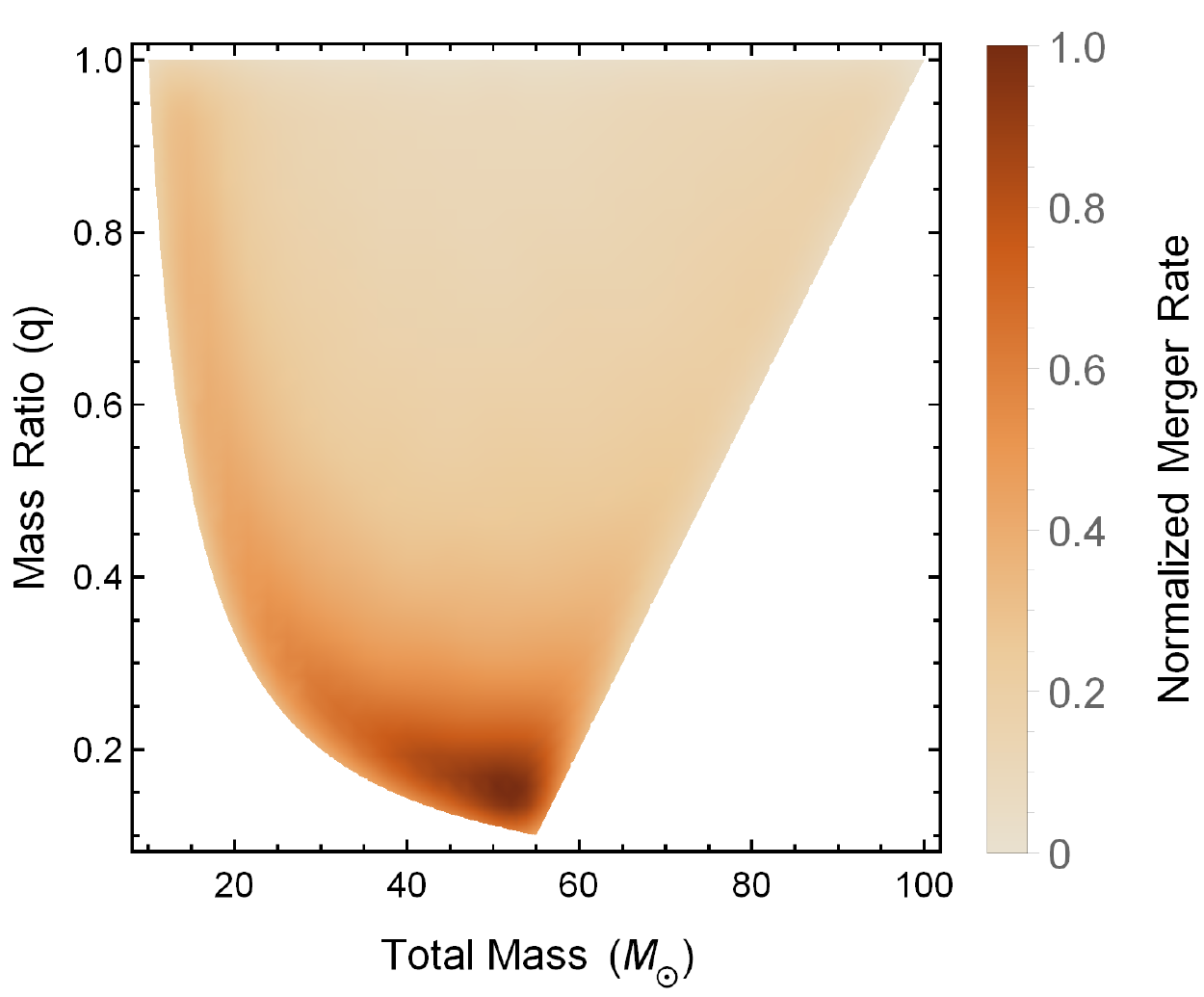}
   \caption{Total binary mass and mass ratio distributions for BBH mergers in AGN disks for SMBH mass $M_\bullet=10^7$\,M$_\odot$, mass distribution index $\beta=2.35$ and accretion rate $\dot{m}_\bullet=0.1$.}
\label{fig:2Dq}
\end{figure}

With the above assumptions the mass ratio distribution is largely independent of the initial mass function index $\beta$ and the SMBH mass, making this distribution universal. Fig. \ref{fig:q} shows the obtained $q$ distributions for different SMBH masses and initial mass functions. We also show in Fig. \ref{fig:2Dq} the 2D mass ratio distribution as a function of total binary mass and mass ratio for the representative case of $M_\bullet=10^7$\,M$_\odot$ and $\beta=2.35$.

In reality, binaries formed in the AGN disk will not represent a random pairing of BHs throughout the disk, as we assumed above. For example as heavier BHs are typically closer to the SMBH, the orbital axis of a heavier BH should typically be closer to another heavier BH than a lower-mass BH, making them preferentially more likely to merge. In this case the resulting mass-ratios will be typically closer to 1 than what we reported above.

\subsection{Connection between AGN accretion rate and merger rate}

\begin{figure}
   \centering  
   \includegraphics[width=0.47\textwidth]{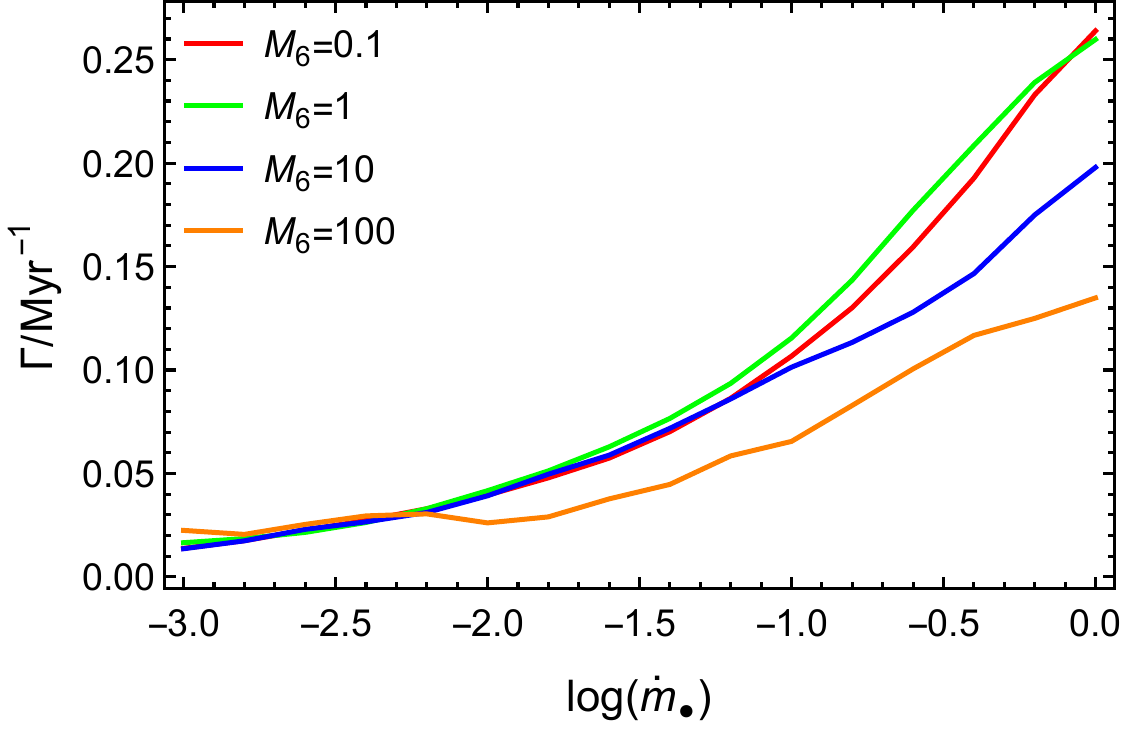}
   \caption{BH merger rate as a function of accretion rate for different SMBH. We study the curves for $M_{\bullet}=10^{5,6,7,8}M_{\odot}$,here we assume $\beta=2.35$.}
\label{fig:accretionrate}   
\end{figure}

In order to calculate the expected merger rate in AGN, we assume that stellar mass in galactic centers are comparable to $2M_{\bullet}$, and that BHs represent 4\% of this mass \citep{2000ApJ...545..847M}. The obtained BBH merger rate as a function of $\dot{m}_{\bullet}$ is shown in Figure \ref{fig:accretionrate} for different SMBH masses. We find that the merger rate is relatively insensitive of the AGN accretion rate: it only changes by about a factor of 10 over three decades of change in the accretion rate from $\dot{m}_{\bullet}=10^{-3}$ to $\dot{m}_{\bullet}=1$. 

The reason for the merger rate's weak dependence on $\dot{m}_{\bullet}$  is that a significant fraction of the BHs in the galactic center|those which are close to alignment at the beginning|take much less time to become fully aligned than $\tau_{\rm agn}$. While the alignment time of these BHs depends on $\dot{m}_{\bullet}$, they will nevertheless merge within $\tau_{\rm agn}$ independently of the accretion rate for the $\dot{m}_{\bullet}$ range we considered. More specifically, we find the approximate proportionality $\tau_{\rm align}\propto \dot{m}_{\bullet}^{0.4}$. This makes the {\it average} merger rate during $\tau_{\rm agn}$ only weakly dependent on $\dot{m}_{\bullet}$. For comparison, the average alignment time for BHs that merge within $\tau_{\rm agn}=10^7$\,yr in an AGN disk with $\dot{m}_\bullet=0.1$ and $M_\bullet=10^7$\,M$_\odot$ is 4.4\,Myr (see also \citealt{1991MNRAS.250..505S,2017ApJ...835..165B}).


This means that the merger rate does not depend strongly on the AGNs' actual accretion rate, therefore knowing the accretion rate for different galaxies does not help differentiate between binary BH formation channels. 

\subsection{Merger rate as a function of SMBH mass}

Since $M_{\bullet}$ affects the AGN disk parameters and the size of the population of BHs close to the disk, the AGN alignment rate will depend on $M_{\bullet}$. We characterized this dependence by calculating the BBH merger rate as a function of $M_{\bullet}$ for the range $M_{\bullet}\in[10^5\mbox{M}_\odot,10^9\mbox{M}_\odot]$. 

\begin{figure}
   \centering  
   \includegraphics[width=0.47\textwidth]{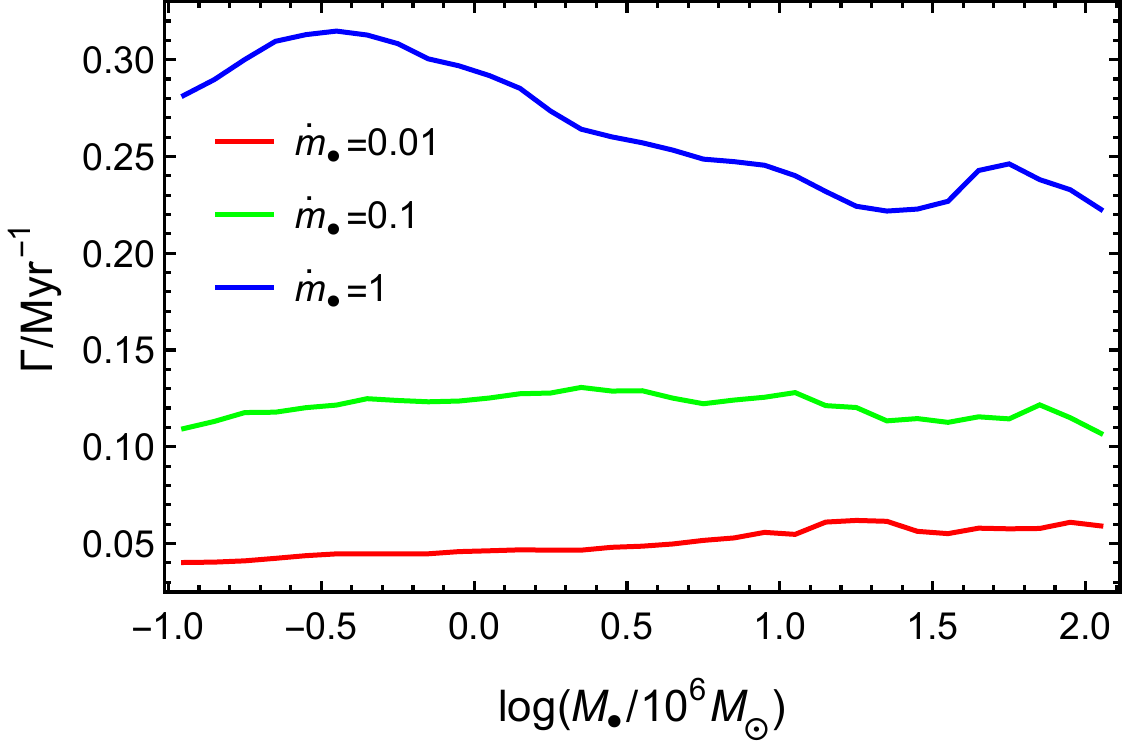}
   \caption{Expected BBH merger rate in an AGN disk as a function of $M_{\bullet}$ for different accretion rates $\dot{m}_{\bullet}$ (see legend). We used $\beta=2.35$.}
\label{fig:rate}   
\end{figure}

Our results are shown in Figure \ref{fig:rate}. We see that the alignment rate continuously, albeit slowly, increases with $M_{\bullet}$. 

\subsection{The overall merger rate in our local universe}

The properties of AGN disks and stellar-mass BH populations in galactic nuclei are not strongly constrained observationally, and the astrophysical parameters determining the merger rate can vary in a broad range \citep{2018ApJ...866...66M}. Therefore, here we aim to give an estimate of the expected BBH merger rate in accretion disks based on our results, using our fiducial parameters.

As we find that low AGN accretion rate does not diminish BH alignment with the disk, we considered the broader group of Seyfert galaxies as opposed to focusing only on highly accreting quasars as was done in previous work \citep{2017ApJ...835..165B}. We adopted a number density of $n_{\rm Seyfert} = 0.018$\,Mpc$^{-3}$ for Seyfert galaxies \citep{2005AJ....129.1795H}. This number density includes galaxies as faint as $M_{\rm r}=-14$. As observations indicate that the SMBH accretion rate for Seyfert galaxies varies within $\dot{m}_{\bullet}=10^{-3}-1$ \citep{2002ApJ...579..530W}, and the merger rate only changes by a factor of few between the two extreme values, for simplicity we adopted a fiducial accretion rate of  $\dot{m}_{\bullet}=0.1$ for all Seyfert galaxies. Further, we considered a fiducial Seyfert SMBH mass of $M_{\bullet}=10^{6}$\,M$_{\odot}$. These choices characterize the Seyfert population and allow us not to rely on the uncertain distribution of these parameters.

For these choices, we find a fiducial overall BBH merger rate of $\sim4$\,Gpc$^{-3}$yr$^{-1}$. With the limited lifetime of AGN disks we assumed here, this merger rate does not appreciably deplete the BH population in galactic nuclei.

The hardening of the BH mass spectrum by AGNs also means that the LIGO-Virgo detection rate of AGN-assisted mergers will be higher than for a similar merger rate with BHs drawn from the initial mass spectrum.  

\section{Star formation in AGN disks}
\label{sec:starformation}

At radii $\gtrsim0.1M_{6}^{1/2}$\,pc from the SMBH, self-gravity becomes important in AGN disks, resulting in fragmentation \citep{2003MNRAS.339..937G}. Some of these fragments will form massive stars, that will in turn produce BH within the AGN disk. Some of these BHs may be born into binaries, which can be hardened to the point of merger by hydrodynamic drag with the surrounding AGN gas \citep{2017MNRAS.464..946S}.  The disk of massive young O and B stars in the central parsec of the Milky Way provides local evidence for a recent episode of disk-mode star formation \citep{2003ApJ...590L..33L}. 

The mass function of stars born from fragments in AGN disks is observed, at least in our own Galactic Center, to have a zero-age main sequence spectral slope of $\sim1.7\pm0.2$, i.e. harder than the Salpeter index of 2.35 \citep{2013ApJ...764..155L}. This top-heavy initial mass function is clearly conducive to compact remnant formation, and all else equal, will produce a shallower BH mass distribution.

However, the mass spectrum of BHs produced in AGN disks will depend both on this harder stellar mass spectrum and stellar metallicity. If higher metallicity stars are produced in AGN disks, this can make the BH mass spectrum significantly steeper \citep{2010ApJ...714.1217B}. 

One important distinction between the {\it in situ} formation channel, and the calculations of \S 3, is the BBH merger rate dependence on $\dot{m}_\bullet$.  Rates of BBH production due to fragmentation of Toomre unstable disks scale linearly with the $\dot{m}_\bullet$ at high accretion rates, and have an even steeper dependence at significantly sub-Eddington accretion rates \citep{2005ApJ...630..167T}.  If BBHs are predominantly formed {\it in situ}, their merger rate $\Gamma$ will thus depend on a steep power of the AGN accretion rate, and the volume-averaged merger rate will be dominated by bright quasars rather than the low-luminosity Seyferts we have found to be of importance for realigning pre-existing BHs into the disk.


\section{Conclusion}
\label{sec:conclusion}

We carried out Monte Carlo simulations of the interaction of stellar-mass BHs and AGN disks to study the effect of these interactions on the properties of BBH mergers. We explored the effect of the initial BH mass function, the mass of the supermassive black hole in the center of the AGN, and the AGN accretion rate. Specifically, we focus on the scenario where singleton stellar-mass BHs from a pre-existing, spherical and isotropic nuclear star clusters are ground down into alignment with the AGN disk, and then efficiently assemble into binaries in disk migration traps \citep{2016ApJ...819L..17B}. Our findings are summarized in the following:
\begin{itemize}
\item The BH mass function hardens due to the mass-dependent impact of the AGN disk. Assuming a power-law initial mass function $M_{\rm bh}^{-\beta}$, the index of the BH mass distribution hardens by $\Delta \beta\approx 1.3$. This change is independent of $\beta$ at least within the $\beta\in[2,3]$ range we tested. The change is also independent of $M_{\bullet}$. This means that independently of the merger distribution in different galaxies, the overall effect will be similar.
\item The BBH merger rate in AGN shows no dependence on the SMBH mass.
\item For our fiducial parameters we find that the BBH merger rate in AGN is $\Gamma\sim4$\,Gpc$^{-3}$yr$^{-1}$, i.e $4-40\%$ of the total BBH merger rate \citep{2018arXiv181112907T,2018arXiv181112940T}. Given the harder mass spectrum, this corresponds to a fraction of $\sim10-50$\% of BBH detections by LIGO/Virgo, assuming that the rest of the detections come from a population of black holes drawn from an initial mass spectrum with $\beta=2.35$.
\item BHs formed within AGN disks may add a comparable number of mergers, further increasing the AGN contribution to the total BBH merger rate \citep{2017MNRAS.464..946S}. This formation channel is strongly correlated with $\dot{m}_{\bullet}$, making it differentiable from BH mergers from disk-alignment. 
\item The merger rate of BBHs produced from BHs aligned by gas drag into AGN disks is only weakly dependent on the SMBH accretion rate. This means that AGNs with lower accretion rates will contribute a substantial part of the overall BBH merger rate from AGNs. This enhances the possibility of finding electromagnetic emission from some BBH mergers as lower-accreting AGNs represent weaker background radiation.
\item Because the merger rate we find has only a weak dependence on $\dot{m}_\bullet$ and $M_\bullet$, the volumetric rate of AGN-aligned BBH mergers will be dominated by the more abundant low-mass Seyferts.  The relatively high volumetric density of these galaxies will necessitate more BBH detections for host galaxy identification techniques than expected if other factors (e.g. {\it in situ} BH formation) we have neglected bias the BBH merger rate in AGNs towards a sub-population of high-$\dot{m}_\bullet$ or high-$M_\bullet$ galaxies \citep{2017NatCo...8..831B}.
\end{itemize}

Future work to refine this analysis will additionally needs to take into account that some BHs are already in binaries before moving into the AGN disk, changing the mass and mass ratio distributions. Spin and the speed of migration within the disk can also affect the final distributions, as can repeated mergers within migration traps \citep{2018arXiv180702859S}. It will also be useful to better understand how AGN-assisted mergers can be differentiated from other dynamical formation channels that also result in hardened mass spectra \citep{2016ApJ...824L..12O}, for example by using their localization \citep{2017NatCo...8..831B,2019arXiv190202797C}.

\begin{acknowledgments}
The authors thank Brian Metzger and Saavik Ford for useful discussions. The authors are thankful for the generous support of the University of Florida and Columbia University in the City of New York. The Columbia Experimental Gravity group is grateful for the generous support of the National Science Foundation under grant PHY-1708028. This work received funding from the European Research Council (ERC) under the European Union's Horizon 2020 research and innovation programme under grant agreement No 638435 (GalNUC) and was supported by the Hungarian National Research, Development, and Innovation Office grant NKFIH KH-125675 (to BK).
\end{acknowledgments}

\bibliography{Refs}

\end{document}